\documentclass[oneside,onecolumn,12pt,a4paper]{article}
\usepackage{graphicx}
\usepackage{dsfont}
\usepackage{natbib}
\usepackage{setspace}
\usepackage{sectsty}
\usepackage{tikz}
\usepackage{amsmath}
\usepackage{hyperref}
\usepackage{color}
\definecolor{darkblue}{rgb}{0.0,0.0,0.3}
\hypersetup{colorlinks,breaklinks,linkcolor=darkblue,urlcolor=darkblue,anchorcolor=darkblue,citecolor=darkblue}

\def\citepos#1{\citeauthor{#1}'s (\citeyear{#1})}

\usepackage{ifthen}
\def\eprinttmp@#1arXiv:#2 [#3]#4@{\ifthenelse{\equal{#3}{}}{\href{http://arxiv.org/abs/#1}{arXiv:#1}}{\href{http://arxiv.org/abs/#2}{arXiv:#2 [#3]}}}
\newcommand{\eprint}[1]{\eprinttmp@#1arXiv: []@}
\newcommand{\doi}[1]{\href{http://dx.doi.org/#1}{doi:#1}}

\begin{document}

\sectionfont{\normalfont\normalsize\bfseries}

\subsectionfont{\normalfont\small\bfseries}

\setstretch{1.4}

\pagenumbering{arabic}
\setcounter{page}{1}

\title{Relativistic Constraints for a Naturalistic Metaphysics of Time}
 \author{\normalsize{Peter W. Evans}\\
 \small{Centre for Time, Department of Philosophy}\\
 \small{The University of Sydney}}
 \maketitle

\begin{abstract}
The traditional metaphysical debate between static and dynamic views in the philosophy of time is examined in light of considerations concerning the nature of time in physical theory. Adapting the formalism of \citet{Rovelli95,Rovelli}, I set out a precise framework in which to characterise the formal structure of time that we find in physical theory. This framework is used to provide a new perspective on the relationship between the metaphysics of time and the special theory of relativity by emphasising the dual representations of time that we find in special relativity. I extend this analysis to the general theory of relativity with a view to prescribing the constraints that must be heeded for a metaphysical theory of time to remain within the bounds of a naturalistic metaphysics.
\end{abstract}

\section{Introduction}

The A-theory of time (often referred to as a \emph{dynamic} view of time), stated briefly, proclaims that temporal passage is an objective feature of reality.\footnote{Of course, there are various A-theoretic views that can be distinguished.} Implicit in this view is that the temporal instant that embodies this passage, the present, maintains a privileged status over and above the temporal instants that have already `passed' (the past) and that are yet to `pass' (the future) and, moreover, this present is in some sense `flowing' through successive instants of time.\footnote{For an illustration of some of the various ways this conception of dynamic time can be expressed, see \citet{Williams51}.} In contrast, the B-theory of time (often referred to as a \emph{static} or \emph{block universe} model of time) is characterised by its rejection of temporal passage as a real and objective feature of the world. As such, there is no privileged instant and all times from the beginning of the universe to the end of the universe are considered to be equally real according to this view.\footnote{The distinction here between A- and B- theories of time follows that of \citet[p.~11]{Dainton}. The A- and B-theories of time can also be characterised as the `tensed' and `tenseless' theories of time, respectively (as in \citet{LePoidevin}, for instance). Under such a construal, the A-theory takes the properties picked out by terms such as past, present and future (known as `tenses') to be real, i.e. to be objective properties of reality. On the other hand, the B-theory denies the reality of tenses. Despite this alternative construal, the core difference between the A- and B-theory remains whether temporal passage is objective or not, as in the above characterisation, and thus `tense' is not taken to be a significant notion here.} The division between these opposing temporal theories defines what we will call the \emph{traditional metaphysical debate} on the nature of time.

It has been suggested that Einstein's special theory of relativity seriously compromises the viability of various formulations of the A-theory of time\footnote{See, for instance, \citet{Rietdijk}, \citet{Putnam}, \citet{Maxwell} and \citet{Saunders}.}; Minkowski's formulation of the special theory of relativity as a four dimensional spacetime has been instrumental in creating the perception that it provides strong evidence for a B-theory of time. On the other hand, much work has been carried out attempting to show the compatibility of special relativity and A-theories of time\footnote{See, for instance, \citet{McCall76}, \citet{Hinchliff}, \citet{Tooley}, \citet{Zimmerman} and \citet{Savitt}.} with a general sentiment emerging that Minkowski spacetime is the wrong sort of entity to definitively adjudicate either way on the traditional debate in the philosophy of time.

This paper is not an attempt to enter this debate and argue for or against either the A- or B-theory of time; nor is it a concern of this paper to attempt to argue the consistency of either of these temporal models with classical relativity theory. The purpose of the current analysis is to investigate and outline the constraints, imposed by the temporal structure of classical physical theory\footnote{Other physical theories, especially quantum theory, may impose further constraints on our temporal models but these will not be considered here.}, that the traditional debate must heed to remain within the bounds of a naturalistic metaphysics.\footnote{A naturalistic metaphysics is meant here \emph{\`{a} la} \citet[Ch.~1]{LadymanRoss}: a ``metaphysics that is motivated exclusively by attempts to unify hypotheses and theories that are taken seriously by contemporary science''.} As one can infer from the introductory remarks above, the special theory of relativity has been conspicuously present in the traditional debate and, therefore, this might make one wonder whether such a project is already \emph{fait accompli}. There are two reasons to be cautious of this presumption. To begin with, existing attempts to answer the question as to why the formal temporal structure of Minkowski spacetime does not preclude the possibility of objective temporal passage (one of which we will meet in \S\ref{sec:proper}) appear to lack a precise characterisation of the picture of time that arises in special relativity. The initial goal of this paper is to adopt a \emph{formal characterisation} of time in special relativity (\S\ref{sec:characterise}) with the resulting picture providing a new perspective on why the constraints imposed by special relativity on the traditional debate are not so restrictive as to quash the debate. The second reason is that the temporal structure of general relativity must be considered also if one is to remain within the bounds of a naturalistic metaphysics. The ultimate goal of this paper, then, is to extend the precise characterisation of time in special relativity to general relativity which, as we will see, imposes much more restrictive constraints.

\subsection{Outline of the paper}

This investigation will proceed as follows. I begin in \S\ref{sec:proper} by sketching an argument from the literature, which I call the proper time argument, to the effect that neither static nor dynamic views of time are precluded solely by the formal temporal structure of Minkowski spacetime. I suggest in \S\ref{sec:characterise} that the proper time argument crucially turns on the dual formulation of time in special relativity and set out a precise framework for characterising time in an attempt so formalise this ambiguity. I employ this formalism to argue for a more general explanation as to why the formal temporal structure of Minkowski spacetime alone precludes neither metaphysical position in the traditional debate. In \S\ref{sec:constrained} I extend the analysis to the picture of time that arises in the general theory of relativity and set out the classical constraints that must be respected by a metaphysical theory of time to remain within the scope of a naturalistic metaphysics.

\section{The proper time argument}
\label{sec:proper}

The most pressing concern for an A-theorist when presented with Minkowski spacetime is the question of how to endow the manifold with an objective temporal passage. For Minkowski spacetime to include temporal passage as an objective element some element contained within the manifold must be in motion. An attractive candidate for this motive element is an objective `now': a hyperplane of simultaneity within spacetime which privileges a particular time instant and which embodies the passage of time. The canonical problem for the A-theorist at this point is that no such hyperplane of simultaneity is privileged as such; due to the relativity of simultaneity, many hyperplanes of simultaneity can be specified depending on the relative motion of the observer and none of these can claim any special status as being a privileged time instant. Thus, it seems as if there is no scope for an objective `now' and thus no scope for objective temporal passage.

While this short argument does indeed provide some important restrictions on the form that an objective temporal passage may take, it does not show that objective temporal passage is incompatible with the formal temporal structure of Minkowski spacetime. It is true that there is no objective hyperplane of simultaneity in Minkowski spacetime and thus no objective \emph{global} `now'. However, a global `now' is not the only candidate for the basis of temporal passage. While an integral element of the special theory of relativity is that there is no absolute fact of the matter about global temporal orderings, there are some facets of Minkowski spacetime that are absolute. The conformal structure of Minkowski spacetime (more on this below) separates the manifold into timelike separated events and spacelike separated events. Observers at the same position in spacetime but in motion relative to one another will define their hyperplane of simultaneity and their local direction of time skewed with respect to one another, but the conformal structure of Minkowski spacetime is inherent in the geometry; they will agree on which regions of spacetime are timelike separated and which regions of spacetime are spacelike separated.

This causal structure of Minkowski spacetime permits that for future directed timelike curves there \emph{is} an objective fact of the matter as to which events are past and which events are future. This temporal ordering of events is only local (i.e. applicable to a single point on a worldline) since observers at different spacetime locations with varied relative motions will disagree on the ordering of spacelike separated, or nonlocal, events. One can then imagine any single spacetime point on a future directed worldline as a candidate for an objective \emph{local} `now'. Minkowski spacetime would then contain many such local objective `nows', each associated with a single worldline. The formal geometric structure of Minkowski spacetime then does \emph{not} preclude the possibility of an objective local `now' (though it certainly does limit the scope of such a `now') and therefore does not preclude out of hand this particular form of objective temporal passage. Let us call this argument the \emph{proper time argument}.\footnote{It is far from obvious that the metaphysical notion of dynamic time that arises from the proper time argument is indeed a viable metaphysical position. The A-theorist who wishes to develop such a view faces a tough challenge explaining exactly how consistency can be maintained between the dynamic local nows to produce the sort of phenomenology that we (as spatiotemporal beings) experience. Indeed, it might seem that relying on the local proper time along a worldline to locate objective temporal passage results in quite a significant modification of the metaphysical position that the A-theorist originally intended. Thus depending upon which features of dynamic time the A-theorist thinks essential, the possibility arises that the metaphysical theory resulting from the above considerations does not do justice to dynamic time. Since this project is not a defence of the A-theory of time, I leave these issues to one side.} I wish to propose here a formal characterisation of the temporal structure of Minkowski spacetime with a view to illustrating precisely why the proper time argument functions as it does.

\section{Characterising time}
\label{sec:characterise}

The proper time argument turns on an ambiguity in the picture of time that arises in the special theory of relativity. There are two ways in which time is formulated in special relativity: the first is as a time measure along an individual worldline, proper time; and the second is as a time measure associated with a coordinatisation of the manifold, coordinate time. Let us consider briefly the details of this ambiguity.\footnote{The exposition here mostly follows \citet{MalamentPP}.}

Minkowski spacetime can be represented by a geometry $(\mathcal{M}^{4},\eta_{\mu\nu})$, which consists of a differentiable, four dimensional manifold, $\mathcal{M}^{4}$, and a flat Lorentzian metric, $\eta_{\mu\nu}$. Given a particular point $p$ in $\mathcal{M}^{4}$, and a four-vector $\mathord{\operatorname{d}}x^{\mu}$ in the tangent space $T_{p}\mathcal{M}^{4}$ at $p$, we can use the metric to construct the line element, $\mathord{\operatorname{d}}s^{2} = \eta_{\mu\nu}\mathord{\operatorname{d}}x^{\mu}\mathord{\operatorname{d}}x^{\nu}$, for each spacetime point in $\mathcal{M}^{4}$. The invariance of $\mathord{\operatorname{d}}s^{2}$ according to the special theory of relativity endows Minkowski spacetime with a \emph{conformal} structure, wherein we can classify $\mathord{\operatorname{d}}s^{2}$ as timelike or spacelike according to its sign. We can then extend these classifications to \emph{curves} in $\mathcal{M}^{4}$, which enables the formalism to model the behaviour of objects in spacetime. We say that a curve is timelike or spacelike if its tangent vector field is characterised as such at every point and thus we can interpret timelike curves as the possible spacetime paths of massive particles; the actual paths of such objects in spacetime are \emph{worldlines}. This provides Minkowski spacetime with a causal structure.

In addition, Minkowski spacetime is temporally orientable: there exists a continuous timelike vector field on $\mathcal{M}^{4}$. We stipulate a temporal orientation to this vector field simply by picking a future direction; any timelike vector at a point of $\mathcal{M}^{4}$ is then \emph{future directed} or \emph{past directed} with respect to this orientation. (As above, a curve is future directed or past directed with respect to this orientation if its tangent vector field is characterised as such at every point.) Time is then associated with the parameter employed to parametrise a future directed timelike curve in $\mathcal{M}^{4}$; such a parametrised curve describes the dynamical behaviour of an object in spacetime (it is only through such a parametrisation that we can begin to speak of `time instants' in special relativity). There are two natural ways that a curve can be parametrised according to an arbitrary observer in spacetime.

Given a future directed timelike curve, $\gamma$, between spacetime points $s_{1}$ and $s_{2}$ in $\mathcal{M}^{4}$ with tangent field $\mathord{\operatorname{d}}x^{\mu}$, we can define the elapsed time between $s_{1}$ and $s_{2}$, $\tau$, with which to parametrise $\gamma$, as the arc length of the curve:
\begin{equation}
  \label{eq:proper}
  \tau = |\gamma| = \int_{s_{1}}^{s_{2}} (\eta_{\mu\nu}\mathord{\operatorname{d}}x^{\mu}\mathord{\operatorname{d}}x^{\nu})^{\frac{1}{2}}~\mathord{\operatorname{d}}s.
\end{equation}
The parametrisation of $\gamma$ by $\tau$ is a `natural' parametrisation since the arc length, as a function of the invariant line element, is a frame independent quantity. We thus call $\tau$ \emph{proper time} and associate it with the time that a clock will measure along its own (not necessarily inertial) worldline.

One can also generate a frame dependent parametrisation of $\gamma$: we can employ clocks at rest with respect to some arbitrary reference frame (proper time along a worldline traced out by an object at rest with respect to this reference frame) to define the elapsed time, $t$, with which to parametrise $\gamma$. By employing this method of parametrisation we have, in effect, stipulated an arbitrary coordinatisation of the manifold, with a time coordinate coinciding with proper time in some arbitrary reference frame, with which to describe spacetime dynamics. We thus call $t$ \emph{coordinate time} and associated it with a global time measure corresponding to the fourth coordinate of the spacetime manifold (so long as the reference frame in question is inertial). Since coordinate time is frame dependent, while proper time is frame independent, the latter is taken to have direct physical significance, while the former is not.\footnote{The distinction between proper time and coordinate time as formulations of time in the special theory of relativity has also been emphasised by \citet{Kroes} and \citet{Rovelli95} and, more recently, by \citet{Savitt}.}

The formal relationship between proper time and coordinate time is given by the Lorentz transformations (which are embodied in $\eta_{\mu\nu}$). Due to this Lorentzian temporal structure, Minkowski spacetime cannot in general be decomposed into distinctly spatial and temporal elements.\footnote{In contrast, recall that the ordinary Euclidean metric imposed on a four dimensional manifold results in a Newtonian spacetime in which space and time can be globally separated as distinct elements of the manifold.} However, provided that one has stipulated a particular time coordinate coinciding with an inertial timelike trajectory, one is able to generate a \emph{foliation} of the Minkowski manifold consisting of spacelike slices orthogonal to the trajectory and thus constituting a set of simultaneous events.

The conformal structure of Minkowski spacetime provides restrictions on how a particular manifold can be foliated. Given these restrictions, however, there remains an infinite number of ways to coordinatise the manifold. With respect to the proper time argument above, this leaves us with no scope to stipulate a global objective `now'. Although objective temporal passage cannot correspond consistently with some objective time coordinate of the manifold, we \emph{are} able to imagine that objective temporal passage corresponds with the incremental evolution along an object's worldline, or the proper time in some reference frame (namely, the reference frame that contains the object in question). Let us consider this variable characterisation of time in special relativity in more formal terms.

In both his \citeyearpar{Rovelli95} and his \citeyearpar{Rovelli}, Rovelli sets about characterising the various roles that the concept of time plays in different scientific theories.\footnote{As well as Rovelli, the different features of time have also been discussed with respect to the special theory of relativity by \citet{Kroes} and with respect to both special and general relativity by \citet{Callender}.} The terminological project associated with this analysis is complicated by the multitude of features that are attributed to the concept of time in natural language. Not often are the entirety of these features found bundled together in the formal structure we identify as time in a physical theory. The following formalism is an adaptation of Rovelli's formalism. Let us begin by considering time as it is often characterised, as a variable $t$ which parametrises the real line $\mathds{R}$.

The real line can be described by the following structure: a manifold, $\mathcal{M}^{1}$, consisting of a set of objects (which in this case is simply all the real numbers) with a one dimensional topology and a differential structure; an ordering, $\mathord{<}$, which sequences the members of the set within the topological structure; a metric, $g$, which ensures that the distance between any two members of the set is meaningfully measurable; and an origin, $\varphi$, which fixes a preferred member of the set. Let us represent this as $\mathds{R}:\{\mathcal{M}^{1},\mathord{<},g,\varphi\}$ \citep[p.~83]{Rovelli95}. It is clear to see that this structure maps into the features we ordinarily associate with the notion of time; the set of objects represent the instants of time, the ordering represents the sequential structure of the instants, the metric represents a measure of temporal duration and the preferred fixed time instant is the present. We should note, however, that this short list of attributes represented by the real line is not a consequence of any particular physical theory. If we consider the picture of time in Newtonian mechanics, for instance, there is no preferred fixed point in the theory that is necessarily labelled as the present. This is not to say that the characterisation of time as the real line is incompatible with Newtonian time; on the contrary, time characterised by the real line is quite consistent with the temporal structure of Newtonian theory. Let us represent the structure of Newtonian time as $N:\{\mathcal{M}^{1},\mathord{<},g\}$ and represent that it is consistent with a richer structure by $N:\{\mathcal{M}^{1},\mathord{<},g\mid\varphi\}$.

The characterisation of time in special relativity is not so straightforward. We saw that the dynamical behaviour of objects in spacetime is described by future directed timelike curves in a four dimensional geometry, $(\mathcal{M}^{4},\eta_{\mu\nu})$, and that the notion of time is associated with the parametrisation of such curves. The significant feature of time in special relativity that sets it apart from Newtonian time is that, for all $p$ in $\mathcal{M}^{4}$, a whole family of future directed timelike curves through $p$ provide a multitude of candidate structures with which time might be identified; the conformal structure of the Minkowski geometry simply does not permit a unique global one dimensional time to be defined in terms of the geometric structure of $\mathcal{M}^{4}$. In other words, it is not possible to define, in terms of the geometric structure of spacetime, a global ordering of all the spacetime points in $\mathcal{M}^{4}$; we can only define a partial ordering on the set of spacetime points, $\mathord{<}^{\prime}$.\footnote{A \emph{total order} on a set S is defined by a binary relation ($\leq$) with the following properties:
\begin{enumerate}
  \item $\forall x \in S$, $x \leq x$,
  \item $\forall x,y \in S$, $x \leq y~\&~y \leq x~\Rightarrow~x=y$,
  \item $\forall x,y,z \in S$, $x\leq y~\&~y \leq z~\Rightarrow~x \leq z$, and
  \item $\forall x,y \in S$, $x \leq y$ or $y \leq x$.
\end{enumerate}
A \emph{partial order} on a set is a binary relation that satisfies (i)-(iii) but not (iv).} There are, however, two avenues open to us for reinstating a total ordering to a set of spacetime points in $\mathcal{M}^{4}$ which correspond to characterising time as coordinate time and proper time, respectively. Let us consider coordinate time first.

While the structure of Minkowski spacetime may not permit a unique global one dimensional time to be directly \emph{definable} from the geometric structure of $\mathcal{M}^{4}$, we are at liberty to \emph{impose} such a structure on the set of all spacetime points. We can simply choose an arbitrary reference frame and take the time as measured by clocks at rest in that frame to provide a unique foliation of the manifold. Of course, a global time measure of this sort is just coordinate time and the unique foliation of $\mathcal{M}^{4}$ into hyperplanes of simultaneity does indeed yield a one dimensional set of time instants (the global hyperplanes), $\mathcal{M}^{1}$, with a total ordering, $\mathord{<}$. A caveat arises at this point, however, when one considers that there is an uncountably infinite number of ways that one can choose such a coordinatisation of the manifold. For every inertial future directed timelike curve through some $p\in\mathcal{M}^{4}$ there is a corresponding foliation of the manifold. Thus there is an infinite number of ways that one might measure the temporal duration between any particular pair of events, corresponding to an infinite number of reference frames, and thus any such measurement in an arbitrary coordinate system is physically meaningless. The characterisation of time in the special theory of relativity as coordinate time thus lacks global \emph{metricity} (i.e. a unique global measure of time). Thus for any reference frame $F$, we can represent coordinate time in special relativity as $C_{S}(F):\{\mathcal{M}^{1},\mathord{<}\}_{F}$.

A second methodology that we can adopt to find a total ordering of a set of spacetime points in $\mathcal{M}^{4}$ is to restrict ourselves to a subset of points in the manifold. Rather than search for a unique \emph{global} one dimensional time, we can instead make use of the linear structure of a single future directed timelike curve to provide a \emph{local} measure of time. Of course, a time measure of this sort is just proper time and the local parametrisation of such a curve yields a one dimensional set of time instants, $\mathcal{M}^{1}$, with a total ordering, $\mathord{<}$. Since proper time is an invariant time measure, the associated parametrisation of a particular worldline is observer independent and thus is a physically meaningful time measure (of temporal durations along the curve only), i.e. proper time is locally metrical. Thus for any timelike curve $\gamma$, we can represent proper time in special relativity as $P_{S}(\gamma):\{\mathcal{M}^{1},\mathord{<},\eta_{\mu\nu}\}_{\gamma}$. In addition, since proper time is only defined locally, fixing a preferred time instant amounts to privileging merely a single spacetime point rather than some global hyperplane. Thus a preferred fixed time instant is consistent with the structure of proper time, $P_{S}(\gamma):\{\mathcal{M}^{1},\mathord{<},\eta_{\mu\nu}\mid\varphi\}_{\gamma}$.

We can also attempt an equally precise construal of dynamic time. We are taking the dynamic view of time here as the claim that we exist in a privileged present that is in some sense `flowing' through successive instants of time. Let us consider which of the above attributes might best fit with this notion of time. Dynamic time is certainly linear, has a well defined order (directed towards the future) and fixes a preferred time instant (the present). Inherent in the idea of `flow' is a notion of continuity that is meaningful only when there exists a measure across the flowing time instants, i.e. dynamic time requires a definite metric. Thus it seems as though dynamic time can be construed as having the structure of the real line as above, $D=\mathds{R}:\{\mathcal{M}^{1},\mathord{<},g,\varphi\}$ (which is hardly a surprise). Let us now reconsider the proper time argument in light of these considerations.

The charge was made against the A-theorist that there can be no objective temporal passage in Minkowski spacetime because there is no scope for an objective hyperplane of simultaneity. This amounts to a claim that not only is there no preferred time instant in special relativity, but a preferred time instant is incompatible with the temporal structure of special relativity. It is clear that this argument aims to characterise time in special relativity as coordinate time and, in light of the above analysis, $C_{S}(F) \neq D$; not only is $C_{S}(F)$ incompatible with a preferred time instant, $C_{S}(F)$ is incompatible with any global and physically meaningful definition of a metric. If the structure of coordinate time were the only formulation of time in special relativity then, to stay within the bounds of a naturalistic metaphysics, dynamic time as we have presented it here would need to be reconsidered as a metaphysical position.

However, we know that time can be construed in special relativity in terms of the structure $P_{S}(\gamma):\{\mathcal{M}^{1},\mathord{<},\eta_{\mu\nu}\mid\varphi\}_{\gamma}$. By formalising the temporal structure of both Minkowski spacetime and the dynamic view of time in this way, we can see immediately that $P_{S}(\gamma)$ is completely consistent with $D:\{\mathcal{M}^{1},\mathord{<},g,\varphi\}$ (given that $\eta_{\mu\nu}$ and $g$ are both `flat' metrics). Thus the dynamic view of time is \emph{not} precluded by the formal temporal structure of Minkowski spacetime. This is then the more compelling explanation as to why the proper time argument functions as it does: the picture of proper time that arises in special relativity ensures that the dynamic view of time is compatible with the temporal structure of Minkowski spacetime due to the correspondence between the characterisations of time that each of them yield. The constraints imposed by the temporal structure of Minkowski spacetime on a naturalistic metaphysics are thus not so restrictive as to force an A-theorist into a major rethink of her position (or a B-theorist, either, for that matter).

\section{The traditional debate constrained}
\label{sec:constrained}

Our description of the dynamical behaviour of objects in spacetime according to the special theory of relativity above is in terms of curves through $\mathcal{M}^{4}$; insofar as this is the case, we are treating spatiotemporal objects as point particles. To provide a more general description of dynamical behaviour in spacetime, we can extend our formalism to treat the general theory of relativity with the addition of \emph{matter fields} to spacetime. A matter field is represented by a smooth tensor field, $T_{\mu\nu}$, on $\mathcal{M}^{4}$ and is assumed to satisfy field equations relating $T_{\mu\nu}$ and the metric. A crucial element to recovering the correspondence between future directed timelike curves on $\mathcal{M}^{4}$ and worldlines of massive particles in spacetime in the special theory of relativity is the latent assumption that the background spacetime structure, $(\mathcal{M}^{4},\eta_{\mu\nu})$, remains fixed \emph{independently} of the $T_{\mu\nu}$ that live on $\mathcal{M}^{4}$.\footnote{\citet[p.~242]{MalamentPP}. We can think of an independent $T_{\mu\nu}$ in the special theory of relativity as representing ``test particles'' in spacetime.} We will call time \emph{independent} when the metric defining time is independent of the matter and energy distribution in the manifold and retain the notation $\eta_{\mu\nu}$ for an independent metric. It is the fact that $T_{\mu\nu}$ is independent of the background spacetime in special relativity that allows us to describe the dynamical behaviour of matter in spacetime in terms of evolution in a time parameter with metric properties (proper time). In contrast when a dependency exists between $T_{\mu\nu}$ and the metric the evolution of the system \emph{defines} proper time and not \emph{vice versa}. General relativity is characterised by such a dependency and we denote the dependent metric $g_{\mu\nu}$.

The geometric structure of general relativity is much the same as the structure we introduced in \S\ref{sec:characterise} for special relativity: we have a geometry $(\mathcal{M}^{4},g_{\mu\nu})$ and we define proper time as before \eqref{eq:proper}. The dependency between $T_{\mu\nu}$ and $g_{\mu\nu}$ is given by the Einstein field equations,
\begin{equation}
  \label{eq:efe}
  G_{\mu\nu}(g_{\mu\nu})=8\pi T_{\mu\nu},
\end{equation}
that define an explicit relation between the matter/energy content of spacetime, represented by the stress-energy tensor $T_{\mu\nu}$, and the curvature of the spacetime manifold, represented by the Einstein tensor $G_{\mu\nu}$ (which is a function of the metric, $g_{\mu\nu}$). Due to this relation, the metric, of which proper time is a function, is a dynamical entity that is at each point in spacetime directly dependent upon the matter/energy density at that point.\footnote{More accurately, there is a \emph{mutual dependency} between the stress-energy tensor and the metric at each point of the manifold.}

The picture of time that arises in general relativity holds similarities with the picture that arises in special relativity; although there are some important differences. Coordinate time can be thought of as an arbitrary foliation of the manifold, each which gives a unique slicing of four dimensional spacetime into a sequence of three dimensional configurations. The linear substructure determined by the foliation yields a total ordering of the slices. However, since the metric is a pointwise function of the matter/energy density of spacetime, it is no longer a flat metric as is the case in special relativity. There is thus no unique notion of parallel transport in curved spacetime and hence there is on way to compare velocities at different points in the manifold unambiguously. Moreover, the parametrisation of the time slices defined by coordinate time in general relativity can be arbitrarily rescaled, which forbids any notion of meaningfully measuring time intervals between pairs of events. This compounds the arbitrariness of such a coordinatisation of the manifold as a temporal measure and so destroys any notion of metricity. Thus for any reference frame $F$, we represent coordinate time in general relativity as $C_{G}(F):\{\mathcal{M}^{1},\mathord{<}\}$. Again, coordinate time is merely an imposition of an arbitrary variable determining time evolution, and because general relativity is foliation invariant, coordinate time has no physical significance.

Proper time in general relativity, on the other hand, is defined exactly as in special relativity \eqref{eq:proper}, except that in general relativity it is determined by a dependent metric, $g_{\mu\nu}$, as above. The local parametrisation of a general relativistic worldline in terms of proper time again yields a one dimensional set of time instants, $\mathcal{M}^{1}$, with a total ordering, $\mathord{<}$. Thus for any timelike curve $\gamma$, we represent proper time in general relativity with the structure $P_{G}(\gamma):\{\mathcal{M}^{1},\mathord{<},g_{\mu\nu}\}_{\gamma}$.

If we now consider the structure of dynamic time, $D$, we can see immediately that similar arguments to those above could be constructed claiming dynamic time to be inconsistent with the structure of general relativity \emph{if} time were characterised simply by $C_{G}(F)$: coordinate time in general relativity has no physically meaningful metric properties. We know, however, not to be persuaded by such argumentation. The case for the consistency of dynamic time with proper time in general relativity, on the other hand, is not so clear cut. We can compare proper time in special relativity, $P_{S}(\gamma):\{\mathcal{M}^{1},\mathord{<},\eta_{\mu\nu}\}_{\gamma}$, to proper time in general relativity, $P_{G}(\gamma):\{\mathcal{M}^{1},\mathord{<},g_{\mu\nu}\}_{\gamma}$, and see that the only difference between the two is the dependency of the metric. The proper time argument of \S\ref{sec:proper} demonstrated that the possibility of locally privileging a temporal instant in a special relativistic spacetime with an independent metric is not prohibited by the formal temporal structure therein. Whether the same can be said for a general relativistic spacetime with a metric that is a pointwise function of the four dimensional matter/energy distribution is the task at hand.

For dynamic time to be consistent with a physical theory there must be a characterisation of time therein that allows us to privilege a present moment that flows objectively. There are two ways that we can understand this privileged present. We can understand the present as ontologically privileged, whereby the notion of flow is envisaged as the existential displacement of the privileged time instant by its successor; accordingly, each time instant then `comes into' existence as the present instant and then `goes out of' existence as a new time instant becomes the present. We can alternatively understand the present as metaphysically privileged, whereby flow is interpreted as the evolution of some property of `presentness' across consecutive time instants that are ontologically undifferentiated. In the special theory of relativity proper time is determined by a fixed background metric structure. The rate of flow of time along a worldline, being determined by the metric, is then not a function of any part of spacetime but the immediate local neighbourhood of the `privileged' instant on the worldline in question; the local flow is determined locally. In this respect, special relativity formally precludes neither an ontologically privileged present nor a metaphysically privileged present, since the flow of time along a worldline does not force us to make an ontological commitment to any part of spacetime but the fixed background structure at a particular spacetime point.

Turning our attention to general relativity, there are two considerations that seem to pull us in opposing directions. The first consideration is that the proper time between two spacetime points on a worldline in general relativity is, just as in special relativity, determined by the metric on the spacetime segment between them, and this is related \emph{locally} (i.e. pointwise) to $T_{\mu\nu}$ via the Einstein field equations. In addition, since it is a principle of general relativity that, for any point of spacetime, we can find a coordinate system in which the metric locally takes the form of the Minkowski metric, there does not appear to be any grounds for a difference between the local properties of time from special relativity to general relativity.

A second consideration, however, suggests that there \emph{is} something global about the ontology of general relativity. \citet*{WestmanSonego} argue that in a generally invariant theory like general relativity, it is untenable to endow a coordinatisation of the manifold, $x^{\mu}$, with operational significance (i.e. as referring to readings on rulers and clocks) since this leads to the underdetermination of Einstein's field equations. Rather, the $x^{\mu}$ must be interpreted merely as mathematical parameters.\footnote{As \citet*[p.~1594]{WestmanSonego} point out, this does not mean that charts on a manifold are arbitrary; rather, the \emph{manifold points} themselves lack operational significance.} This amounts to the claim that $\mathcal{M}^{4}$ cannot represent something empirically accessible in general relativity. What is empirically accessible, according to \citet*{WestmanSonego}, is the coincident values of different measurable physical quantities (field values) that motivate a refined notion of an event, which they label a ``point-coincidence''. The set of all point-coincidences, which possesses a natural manifold structure, denoted $\mathcal{E}$, turns out to be a natural representation of the totality of physical events (i.e. spacetime). In such a representation $\mathcal{M}^{4}$ plays no empirical role and only the mutual relationships of the configurations of various fields are physically relevant. Thus, the suggestion is that general relativity must be interpreted as having a kind of \emph{relational} ontology.

It is hard, then, to envisage an ontologically privileged present in a general relativistic spacetime given the portent here that general relativity is predicated upon a coordinate invariant notion of ontology. The exclusivity of the reality of a locally defined present time instant (as required by an ontologically privileged present) seems to be compromised by the relational nature of the ontology of general relativity. Thus one might struggle to justify a metaphysical theory of classical time, which remains within the scope of a naturalistic metaphysics, and that interprets flow in terms of the existential displacement of a privileged time instant by its successor. To remain within these constraints, the traditional debate must proceed in the following manner: if one wanted to maintain that there is an objectively flowing privileged time instant, then one must understand this instant to be metaphysically privileged, whereby flow is interpreted as the evolution of the property `presentness' across consecutive time instants that are ontologically undifferentiated.

What is far from obvious is whether this picture yields a nontrivial metaphysical theory of time; for instance, in what meaningful sense is this conception of the present `privileged' or `objective', especially if we are simply positing a preferred temporal instant with this property to allow us to maintain that we occupy an A-theoretic reality? Whether or not there remains logical space for an A-theory of time within these constraints depends upon the way in which the A-theorist wishes to refine the notion of the privileged present.\footnote{\citet{Zimmerman10} sets out a comprehensive defence of an A-theory of time that fits explicitly within these constraints.} As an integral element to any ensuing analysis here, I wish to point out in a sceptical spirit that the dynamic view of time seems to be beset by the imprecise and obscure nature of notions such as `privileged', `objective' and `flow' and it is not entirely clear that these terms are conducive to rigorous definition in this context.\footnote{Though see \citet{Price09} for a recent (and not very sympathetic) analysis of flow.} The implication, then, is that the A-theorist who respects naturalistic metaphysics owes us an account of the dynamic view of time that avoids the triviality of merely stipulating a spacetime point as objectively metaphysically distinguished.

There is a further caveat that jeopardises the viability of a dynamic view of time.\footnote{The formalism here follows \citet{Belot}} Even if we consider that each individual worldline in spacetime is a vehicle of objective flow, to ensure that every such worldline yields a totally ordered linear subset of the manifold we require the existence of a spacelike hypersurface $\Sigma\subset\mathcal{M}^{4}$ with the property that every inextendible timelike curve in $\mathcal{M}^{4}$ intersects $\Sigma$ exactly once. We call $\Sigma$ a \emph{Cauchy surface} and note that it follows from this condition that $\Sigma$ is a three dimensional spacelike submanifold of $\mathcal{M}^{4}$. A geometry $(\mathcal{M}^{4},g_{\mu\nu})$ that admits the existence of a Cauchy surface is said to be \emph{globally hyperbolic}. If $(\mathcal{M}^{4},g_{\mu\nu})$ is globally hyperbolic then $\mathcal{M}^{4}$ is diffeomorphic to a manifold of the form $\Sigma\times\mathds{R}$ (where we take $\Sigma$ here to represent a diffeomorphism equivalence class of three dimensional Cauchy surfaces) \citep{Geroch}. Thus a necessary condition for the possibility of dynamic time is the requirement that our reality be represented by a manifold $\mathcal{M}^{4}$ that can be foliated by Cauchy surfaces.\footnote{Of course, one could argue that a total ordering of temporal instants is not essential to the dynamic view of time, i.e. dynamic time might be better characterised by the structure $D^{\prime}:(\mathcal{M}^{1},\mathord{<}^{\prime},g,\varphi)$, with a partial ordering $\mathord{<}^{\prime}$. Global hyperbolicity would not be a necessary condition for the possibility of a dynamic time represented by $D^{\prime}$.} A problem arises here for the dynamic view of time since only a subset of the solutions to Einstein's field equations \eqref{eq:efe} have this property; \citepos{Goedel} infamous and eponymous spacetime solution, which contains closed timelike curves, is just one example of a solution to the field equations that is not globally hyperbolic.\footnote{On the other hand, just because a physical theory admits the \emph{possibility} of a solution to the field equations of a particular sort does not imply that this solution is \emph{necessarily} physically realisable: think of the case of a pendulum with negative length. More to the point, there are as yet no solutions to the field equations that have been used to make physical predictions that are not globally hyperbolic. Such argumentation may alleviate the worry an A-theorist might have with G\"{o}del-type universes in the first place.}

There is, however, a potential reprieve for the A-theorist in this case. The set of spacetime solutions to Einstein's field equations that can be foliated into spacelike hypersurfaces have taken on considerable significance over the last half a century.\footnote{See \citet{Dirac58}, \citet{Bergmann} and \citet*{Arnowitt}.} The restriction to globally hyperbolic spacetime solutions is required for the Hamiltonian formulation of general relativity and this in turn is integral to using canonical quantisation techniques to develop a quantum theory of gravity. Thus, it may turn out that a successful quantum theory of gravity provides independent evidence that our spacetime is indeed globally hyperbolic, thus admitting the existence of Cauchy surfaces. This would ensure that each individual worldline in spacetime consisted of a totally ordered linear subset of the manifold, which would rekindle the possibility that each worldline is a vehicle of objective flow.

The catch, however, is that this reprieve is only plausible if it is possible to find a physical basis for fixing a preferred foliation of the spacetime manifold, which is a difficult task to say the least for a foliation invariant theory such as general relativity. A suggestion has been made in recent times, however, that the so-called \emph{constant mean curvature} (CMC) foliation approach provides just this: a unique foliation for a reasonably large subset of spacetime solutions, which are determined by constraining the possible ways that $\Sigma$ is permitted to be embedded in $\mathcal{M}^{4}$.\footnote{See \citet{Wuethrich}.} This is achieved by expressing the content of the Einstein field equations in terms of the Hamiltonian constraint equations that we obtain when the canonical variables are the 3-metric and extrinsic curvature of $\Sigma$. We can then define a subset of the spacetime solutions by the condition that the mean of the extrinsic curvature is constant across $\Sigma$. As it happens, parametrising the hypersurfaces of a spacetime by constant mean curvature leads to a unique foliation of the spacetime. The A-theorists hopes for a potential reprieve, then, are pinned to whether or not a physical basis for privileging CMC foliation can be found.

To conclude this paper I wish to briefly remark on two significant issues that foreshadow the A-theorists program in connection to the privileged foliation issue. On the bright side for the A-theorist, \citet{Belot} argues that the CMC foliation approach may be an instrumental ingredient in solving the \emph{problem of time} in general relativity. However, he also concedes that the approach ``violates the spirit of general relativity'' in that it reinstates a privileged distinction between time and space \citeyearpar[p.~219]{Belot}. On the not so bright side for the A-theorist, \citet{Wuethrich} sets out a rather comprehensive and convincing argument against the possibility of using the CMC approach to support a particular A-theory of time: \emph{presentism}. Thus while it seems that the A-theorist \emph{may} find supporting physical structure in the Hamiltonian formulation of general relativity, there are significant obstacles still to be overcome.

\end{document}